\documentclass[12pt]{article}
\usepackage{amsmath}
\usepackage{amssymb}
\begin{document}
\vskip 1truein
\begin{center}
{\bf \Large{On unitarity of a Yang-Mills type formulation for
massless and massive gravity with propagating torsion}}
 \vskip 5pt
{Rolando Gaitan D.}${}^{a
}${\footnote {e-mail: rgaitan@uc.edu.ve}} \\
${}^a${\it Grupo de F\'\i sica Te\'orica, Departamento de F\'\i
sica, Facultad de Ciencias y Tecnolog\'\i a, Universidad de
Carabobo, A.P. 129
Valencia 2001, Edo. Carabobo, Venezuela.}\\
\end{center}
\vskip .2truein
\begin{abstract}

A perturbative regime based on contorsion as a dynamical variable
and metric as a (classical) fixed background, is performed in the
context of a pure Yang-Mills formulation based on $GL(3,R)$ gauge
group. In the massless case we show that the theory propagates
three degrees of freedom and only one is a non-unitary mode. Next,
we introduce quadratical terms dependent on torsion, which
preserve parity and general covariance. The linearized version
reproduces an analogue Hilbert-Einstein-Fierz-Pauli unitary
massive theory plus three massless modes, two of them non-unitary
ones. Finally we confirm the existence of a family of unitary
Yang-Mills-extended theories which are classically consistent with
Einstein's solutions coming from non massive and topologically
massive gravity.

\end{abstract}

\vskip .2truein
\section{Introduction}

There were some contributions on the exploration of classical
consistency of a pure Yang-Mills (YM) type formulation for
gravity, including the cosmological extension \cite{g1,g2} (and
the references therein), among others. In those references,
Einstein's theory is recovered after the imposition of torsion
constraints.

Unfortunately, the path to a quantum version (if it is finally
possible) is not straightforward. For example, it is well known
that the Lagrangian of a pure YM theory based on the Lorentz group
$SO(3,1)\simeq SL(2,C)$ \cite{kp} leads to a non-positive
Hamiltonian (due to non-compactness of the aforementioned gauge
group) and, then the canonical quantization procedure fails.
However, there is a possible way out if it is considered an
extension of the YM model thinking about a theory like
Gauss-Bonnet with Torsion\cite{kp} and this is confirmed because
the existence of a possible family of quadratical curvature
theories from which can be recovered unitarity\cite{sez}.

A first aim of this work is to expose, with some detail, a similar
(and obvious) situation about non-unitarity in a YM formulation
with $GL(3,R)$ as a gauge group in both massless and massive
theories. There is an interest focussed in the study of massive
gravity and propagating torsion\cite{hvn}, among others.
Particularly, the massive versions that we shall explore here
arise, on one hand from some quadratical terms set ($T^2$-terms)
preserving parity which depends on torsion (the old idea about
considering $T^2$-terms in a dynamical theory of torsion has been
considered in the past\cite{hs}) and, at a perturbative regime
they give rise to a Fierz-Pauli's massive term. On the other hand,
we review the topologically masive version of the YM
gravity\cite{g2} which do not preserves parity and how is the way
to reach unitarity.

Whatever the model considered, throughout this work we follow the
spirit of Kibble's idea\cite{kibble} treating the metric as a
fixed background, meanwhile the torsion (contorsion) shall be
considered as a dynamical field and it would be thought as a
quantum fluctuation around a classical fixed background.

This paper is organized as follows. The next section is devoted to
a brief review on notation of the cosmologically extended YM
formulation\cite{g1} in $N$-dimensions and its topologically
massive version in $2+1$ dimension\cite{g2}. In section 3, we
consider the scheme of linearization of the massless theory around
a fixed Minkowskian background, allowing fluctuations on torsion.
Next, the Lagrangian analysis of constraints and construction of
the reduced action is performed, showing that this theory does
propagate degrees of freedom, including a ghost. In section 4, we
introduce an appropiate $T^2$-terms, which preserve parity,
general covariance, and its linearization gives rise to a
Fierz-Pauli mass term. There, the non-positive definite
Hamiltonian problem gets worse: the Lagrangian analysis shows that
the theory has more non-unitary degrees of freedom and we can't
expect other thing. Gauge transformations are explored in section
5. Although $T^2$-terms provide mass only to some spin component
of contorsion, the linearized theory loses the gauge invariance
and there is no residual invariance. This is clearly established
through a standard procedure for the study of possible chains of
gauge generators\cite{c}. In section 6 we confirm the well known
fact that there exists a family of theories which can cure the
ghost problem\cite{sez} and they are classically consistent when
it is shown that the set of solutions contains the Einsteinians
ones. We end up with some concluding remarks.

\vskip .2truein
\section{A pure Yang-Mills formulation for gravity: massless and topological massive cases}

Let $M$ be an $N$-dimensional manifold with a metric, $g_{\mu \nu
}$ provided. A (principal) fiber bundle is constructed with $M$
and a 1-form connection is given,  ${(A_\lambda)^\mu}_\nu $ which
will be non metric dependent. The affine connection transforms as
${A_\lambda}^\prime =UA_\lambda U^{-1} + U
\partial _\lambda U^{-1}$ under $U \in GL(N,R)$. Torsion and curvature tensors are
${T^\mu}_{\lambda\nu}={(A_\lambda)^\mu}_\nu-{(A_\nu)^\mu}_\lambda$
and $F_{\mu\nu} \equiv \partial_\mu A_\nu - \partial_\nu A_\mu +
[A_\mu , A_\nu ]$, respectively. Components of the Riemann tensor
are ${R^\sigma}_{\alpha \mu \nu }\equiv ({F_{\nu \mu})^\sigma
}_\alpha$. The gauge invariant action with cosmological
contribution is\cite{g1}
\begin{equation}
{S^{(N)}}_0=\kappa^{2(4-N)} \big<-\frac 14 \, tr\, F^{\alpha
\beta}F_{\alpha \beta}+ q(N) \lambda^2 \big> \,\, , \label{eqa1}
\end{equation}
where $\kappa^2$ is in length units, $\big<...\big>\equiv \int d^N
x \sqrt{-g}(...) $, $\lambda$ is the cosmologic constant and the
parameter  $q(N)=2(4-N)/(N-2)^2(N-1)$ depends on dimension. The
shape of  $q(N)$ allows the recovering of (free) Einstein's
equations as a particular solution when the torsionless Lagrangian
constraints are imposed and $q(N)$ changes it sign when $N>5$. The
field equations are ${T_g}^{\alpha\beta}=
-\kappa^2g^{\alpha\beta}\lambda^2$ where
${T_g}^{\alpha\beta}\equiv
\kappa^2\,tr[F^{\alpha\sigma}{F^\beta}_\sigma
-\frac{g^{\alpha\beta}}4 \, F^{\mu \nu}F_{\mu \nu}]$ is the
energy-momentum tensor of gravity, and equation coming from
variation of connection is $\frac 1{\sqrt{-g}}\,\,\partial _\alpha
(\sqrt{-g}\,\,F^{\alpha \lambda}) + [A_\alpha , F^{\alpha
\lambda}] =0$, which can be rewritten as follows
\begin{eqnarray}
\nabla_\mu R_{\sigma\lambda}-\nabla_\lambda R_{\sigma\mu} =0\,\,
,\,\, \label{g1a}
\end{eqnarray}
and the trace $\sigma-\lambda$ gives the expected condition
$R=constant$.

It is well known that the introduction of a Chern-Simons
lagrangian term (CS) in the Hilbert-Einstein formulation of
gravity provides a theory which describes a massive excitation of
a graviton in 2+1 dimensions\cite{DJT}. If a cosmological term is
included, the cosmologically extended topological massive gravity
(TMG$\lambda$) arises\cite{d}. The aforementioned action is
\begin{eqnarray}
S=\frac{1}{\kappa^2} \int d^3 x \sqrt{-g}(R+\lambda)+
\frac{1}{\kappa^2\mu}\,S_{CS}\,\, \,, \label{eqt1}
\end{eqnarray}
where $\mu$  is in $(lenght)^{-1}$ units and $S_{CS}$ is the CS
action. In a Riemannian space-time, the action (\ref{eqt1}) gives
the field equation, $R^{\mu \nu }-\frac{g^{\mu \nu}}2R-\lambda
g^{\mu \nu}+\frac{1}{\mu}\,C^{\mu \nu }=0$ where $C^{\mu \nu}$ is
the (traceless) Cotton tensor. The trace of the field equation
gives a consistency condition on the trace of the Ricci tensor
(this means, $R=-6\lambda$). Starting with the field equation, it
is possible to write down an hyperbolic-causal equation which
describes a massive propagation for the Ricci tensor as follows
\begin{eqnarray}
(\nabla_\mu \nabla^\mu -
\mu^2)\,R_{\mu\nu}-R^{\alpha\beta}R_{\alpha\beta}g_{\mu\nu}
+3{R^\alpha }_\mu R_{\alpha\nu} +\frac{\mu^2}{3}\,Rg_{\mu\nu}
\nonumber \\ -\frac{3}{2}\,RR_{\mu\nu} +\frac{1}{2}\, R^2
g_{\mu\nu} =0
 \, \, , \label{eqt2bb}
\end{eqnarray}
where $\nabla_\mu$ is taken with Christoffel's symbols.

Next, we can explore consistence of a Yang-Mills type formulation
for topological massive gravity with cosmological constant
(GTMG$\lambda$), verifying the existence of causal propagation and
the fact that standard TMG$\lambda$ can be recovered from
GTMG$\lambda$ at the torsionless limit. The GTMG$\lambda$ model
is\cite{g2}
\begin{equation}
S_{GTMG\lambda}={S^{(3)}}_0 +
\frac{m\kappa^2}{2}\big<\varepsilon^{\mu\nu\lambda}\,tr\big( A_\mu
\partial_\nu A_\lambda +\frac{2}{3}\, A_\mu A_\nu A_\lambda \big)\big>\,
\, , \label{eqa2}
\end{equation}
which does not preserve parity and ${S^{(3)}}_0$ is given by
(\ref{eqa1}) for $N=3$. Moreover this model is gauge variant
because the Chern-Simons transforms like
\begin{equation}
\delta_U S_{CS}=-\frac{m\kappa^2}{2}\int
d^3x\,\epsilon^{\mu\nu\lambda}\,tr\,\partial_\nu \big[ A_\mu
\partial_\lambda U U^{-1}\big]-4\pi^2\kappa^2m\,W(U)
 \, \, , \label{ssm15}
\end{equation}
where $W(U)\equiv \frac{1}{24\pi^2}\int
d^3x\,\epsilon^{\mu\nu\lambda}\,tr\big( U^{-1}\partial_{\mu}
UU^{-1}\partial_{\nu} UU^{-1}\partial_{\lambda} U\big)$ is the
''winding number'' of the gauge transformation  $U$.

The torsionless limit of (\ref{eqa2}) can be explored by
introducing nine torsion's constraints through the new action $S'
= S_{GTMG\lambda}+\kappa^2\int d^3 x \sqrt{-g}
\,b_{\alpha\beta}\,\varepsilon^{\beta\lambda\sigma}{(A_\lambda)^\alpha}_\sigma$,
where $b_{\alpha\beta}$ are Lagrange multipliers. Variation on
connection and metric gives rise the following field equations
\begin{equation}
\nabla_\mu R_{\sigma\lambda}-\nabla_\lambda R_{\sigma\mu}
-m\,{\varepsilon^{\nu\rho}}_\sigma(g_{\lambda\nu}R_{\mu\rho}-g_{\mu\nu}R_{\lambda\rho}
-\frac{2}{3}\,Rg_{\lambda\nu}g_{\mu\rho})=0\,\, , \label{eq13}
\end{equation}
\begin{equation}
R_{\sigma\mu}{R^\sigma}_\nu
-RR_{\mu\nu}+\frac{g_{\mu\nu}}{4}\,R^2-g_{\mu\nu}\lambda^2=0\,\, ,
\label{eqa14}
\end{equation}
and Lagrange multipliers are
\begin{equation}
b_{\mu\nu}=\frac{mR}{6}\,g_{\mu\nu}\,\, . \label{eqa14a}
\end{equation}

The trace $\sigma-\lambda$ of (\ref{eq13}) leads to the following
consistency condition
\begin{equation}
R=constant \,\, , \label{eqam10}
\end{equation}
and due to this condition on the Ricci scalar, we can test
particular solutions of the type
$R_{\mu\nu}=\frac{R}{3}\,g_{\mu\nu}$, by pluging them in
(\ref{eqa14}), and this gives
\begin{eqnarray}
R=\pm 6\mid\lambda\mid\,\, , \label{eqam13}
\end{eqnarray}
verifying the existence of (Anti) de Sitter solutions.

A quick look on causal propagation of the theory can be performed
writing a second order equation from (\ref{eq13}), this means
\begin{eqnarray}
(\nabla_\alpha \nabla^\alpha -
m^2)\,R_{\mu\nu}-R^{\alpha\beta}R_{\alpha\beta}g_{\mu\nu}
+3{R^\alpha }_\mu R_{\alpha\nu} +\frac{m^2R}{3}\,g_{\mu\nu}
\nonumber \\-\frac{3R}{2}\,R_{\mu\nu} +\frac{R^2}{2}\,  g_{\mu\nu}
=0
 \, \, , \label{eq4}
\end{eqnarray}
which describes a massive hyperbolic-causal propagation of
graviton. So, GTMG$\lambda$ contains as a particular case the
TMG$\lambda$ classical formulation (at the torsionless limit) if
we take the mass value $m$ as the CS ($m=\mu$) and the consistency
condition (\ref{eqam10}) is fixed as (\ref{eqam13}).

We underline that GTMG$\lambda$ is gauge variant under $GL(3,R)$,
due to the presence of the CS term. However, it is well known by
taking boundary conditions on the elements $U$, the term $\int
d^3x\,\epsilon^{\mu\nu\lambda}\,tr\,\partial_\nu \big[ A_\mu
\partial_\lambda U U^{-1}\big]$ in (\ref{ssm15}), goes to zero
and the transformation rule now is $\delta_U
S=-4\pi^2\kappa^2m\,W(U)$. If we demand that the expectation value
of a gauge invariant operator (i.e., $<\mathcal{O}>\equiv
Z^{-1}\int \mathcal{D}A \,\mathcal{O}(A) \,e^{iS}$ with the gauge
invariant measure $\mathcal{D}A$ and the normalization constant
$Z$) must be gauge invariant too, it is required that
$-4\pi^2\kappa^2m\,W(U)$ be an integral multiple of $2\pi$ and a
quantization condition on the parameter $\kappa^2m$ must arises.
This fact occurs, at least, by performing a restriction on the
covariance of the theory, this means, taking a compact subgroup of
$GL(3,R)$ (i.e., $SO(3)$).

\vskip .2truein
\section{Linearization of the massless theory}
With a view on the performing of a perturbative study of the
massive model, we wish to note some aspects of the variational
analysis of free action (\ref{eqa1}) in $2+1$ dimensions. As we
had said above, the connection shall be considered as  a dynamical
field whereas the space-time metric would be a fixed background,
in order to explore (in some sense) the isolated behavior of
torsion (contorsion) and avoid higher order terms in the field
equations. For simplicity we shall assume $\lambda =0$.

Then, let us  consider a Minkowskian space-time with a metric
$diag(-1,1,1)$ provided and, obviously with no curvature nor
torsion. The notation is
\begin{equation}
\overline{g}_{\alpha\beta}=\eta_{\alpha\beta}\, \, , \label{eqb1}
\end{equation}
\begin{equation}
{\overline{F}}^{\alpha\beta}=0\, \, , \label{eqb2}
\end{equation}
\begin{equation}
{\overline{T}^\lambda}_{\mu\nu}=0\, \, . \label{eqb3}
\end{equation}
It can be observed that curvature ${\overline{F}}^{\alpha\beta}=0$
and torsion ${\overline{T}^\lambda}_{\mu\nu}=0$, in a space-time
with metric $\overline{g}_{\alpha\beta}=\eta_{\alpha\beta}$
satisfy the background equations, $\frac
1{\sqrt{-\overline{g}}}\,\,\partial _\alpha
(\sqrt{-\overline{g}}\,\,\overline{F}^{\alpha \lambda}) +
[\overline{A}_\alpha , \overline{F}^{\alpha \lambda}] =0$ and
${\overline{T}_g}^{\alpha\beta}=0$, identically.

Thinking in variations
\begin{equation}
A_\mu=\overline{A}_\mu+ a_\mu\,\,\,\,,\,\,\,\,|a_\mu|\ll 1\, \, ,
\label{eqb4}
\end{equation}
for this case $\overline{A}_\mu=0$. Then,  action (\ref{eqa1})
takes the form
\begin{equation}
{S^{(3)L}}_0 =\kappa^2 \big<-\frac 14 \, tr\, f^{\alpha
\beta}(a)f_{\alpha \beta}(a)\big> \,\, , \label{eqb5}
\end{equation}
where $f_{\alpha \beta}(a)=\partial_\alpha a_\beta-\partial_\beta
a_\alpha $ and (\ref{eqb5}) is gauge invariant under
\begin{equation}
\delta a_\mu = \partial_\mu \omega  \,\, , \label{eqb6}
\end{equation}
with
$\omega \in G=U(1)\times...{3^2}...\times U(1)$.

In order to describe in detail the action (\ref{eqb5}), let us
consider the following decomposition for perturbed connection
\begin{equation}
{(a_\mu)^{\alpha}}_\beta={\epsilon^{\sigma\alpha}}_\beta
k_{\mu\sigma}+{\delta^\alpha}_\mu v_\beta
-\eta_{\mu\beta}v^\alpha\,\, , \label{eqb7}
\end{equation}
where $k_{\mu\nu}=k_{\nu\mu}$ and $v_\mu$ are the symmetric and
antisymmetric parts of the rank two perturbed contorsion  (i. e.,
the rank two contorsion is $K_{\mu\nu}\equiv -\frac{1}2
{\epsilon^{\sigma\rho}}_\nu K_{\sigma\mu\rho}$), respectively. It
can be noted that decomposition  (\ref{eqb7}) has not been
performed in irreducible spin components and explicit writing down
of the traceless part of $k_{\mu\nu}$ would be needed. This
component will be considered when the study of reduced action
shall be performed. Using (\ref{eqb7}) in (\ref{eqb5}), we get
\begin{equation}
{S^{(3)L}}_0 =\kappa^2 \big< k_{\mu\nu}\Box{}k^{\mu\nu}+
\partial_\mu k^{\mu\sigma}\partial_\nu {k^\nu}_\sigma
-2 \epsilon^{\sigma\alpha\beta}\partial_\alpha v_\beta\partial_\nu
{k^\nu}_\sigma - v_\mu \Box{}v^\mu +(\partial_\mu v^\mu)^2\big>
\,\, , \label{eqb8}
\end{equation}
which is gauge invariant under the following transformation rules
(induced by (\ref{eqb6}))
\begin{equation}
\delta k_{\mu\nu}= \partial_\mu \xi_\nu + \partial_\nu \xi_\mu
\,\, , \label{eqb9}
\end{equation}
\begin{equation}
\delta v_\mu =-{\epsilon^{\sigma\rho}}_\mu \partial_\sigma
\xi_\rho \,\, , \label{eqb10}
\end{equation}
with $\xi_\mu \equiv \frac{1}4
{\epsilon^\beta}_{\alpha\mu}{w^\alpha}_\beta$. These
transformation rules clearly show that only the antisymmetric part
of $w$ is needed (i. e.: only three gauge fixation would be
chosen).

In expression (\ref{eqb8}) we can observe that the term $v_\mu
\Box{}v^\mu$ has a wrong sign, telling us about the non-unitarity
property of the theory. However, field equations are
\begin{equation}
2\Box{}k_{\mu\nu}-
\partial_\mu\partial_\sigma {k^\sigma}_\nu-\partial_\nu\partial_\sigma {k^\sigma}_\mu
+{\epsilon^{\sigma\rho}}_\mu\partial_\nu\partial_\sigma v_\rho
+{\epsilon^{\sigma\rho}}_\nu\partial_\mu\partial_\sigma v_\rho =0
\,\, , \label{eqb11}
\end{equation}
\begin{equation}
\epsilon^{\sigma\rho\beta}\partial_\sigma \partial_\mu
{k^\mu}_\rho +\Box{}v^\beta +\partial^\beta\partial_\mu v^\mu=0
\,\, , \label{eqb12}
\end{equation}
and note that  (\ref{eqb11}) satisfies the consistency condition
\begin{equation}
\Box{}k-\partial_\mu\partial_\nu k^{\mu\nu}=0 \,\,. \label{eqb12a}
\end{equation}

Divergence of (\ref{eqb12}) says that $\partial_\mu v^\mu$ is a
massless 0-form then, if we define $\hat{\partial}_\sigma\equiv
\Box{}^{-\frac{1}2}\partial_\sigma $, the following relation can
be written
\begin{equation}
v^\beta=-\epsilon^{\sigma\rho\beta}\hat{\partial}_\sigma
\hat{\partial}_\mu {k^\mu}_\rho  \,\, , \label{eqb13}
\end{equation}
up to a massless-transverse 1-form. Using  (\ref{eqb13}) in
(\ref{eqb11}), gives rise to
\begin{equation}
\Box{}k_{\mu\nu}-
\partial_\mu\partial_\sigma {k^\sigma}_\nu-\partial_\nu\partial_\sigma {k^\sigma}_\mu
+\partial_\mu\partial_\nu k=0 \,\, , \label{eqb14}
\end{equation}
up to a massless 0-form. This last equation with condition
(\ref{eqb12a}) would suggests a possible equivalence with the
model for gravitons of the linearized Hilbert-Einstein theory (i.
e.: free gravity in $2+1$ does not propagate degrees of freedom).
However, this suggestion is wrong because we were dropped out some
light modes and, then it is necessary to take into account both
massive and massless complete sets of modes (light modes are
relevant at the lower energy regime).

Now, let us study the system of Lagrangian constraints in order to
explore the number of degrees of freedom. A possible approach
consists in a $2+1$ decomposition of the action (\ref{eqb8}) in
the way
\begin{eqnarray}
{S^{(3)L}}_{0} =\kappa^2
\big<[-\dot{k}_{0i}+2\partial_ik_{00}-2\partial_nk_{ni}-2\epsilon_{in}\dot{v}_n
+2\epsilon_{in}\partial_n v_0 ]\dot{k}_{0i}\nonumber
\\+\dot{k}_{ij}\dot{k}_{ij}
+[2\epsilon_{nj}\partial_nk_{00}+2\epsilon_{nj}\partial_mk_{nm}-\dot{v}_j-2\partial_jv_0]\dot{v}_j\nonumber
\\+2(\dot{v}_0)^2 +k_{00}\Delta k_{00}-2k_{0i}\Delta
k_{0i}+k_{ij}\Delta k_{ij}-(\partial_ik_{i0})^2\nonumber
\\+\partial_nk_{ni}\partial_mk_{mi} -2\epsilon_{ij}\partial_i v_j
\partial_nk_{n0}-2\epsilon_{lm}\partial_m v_0
\partial_nk_{nl}+\nonumber \\v_0\Delta v_0-v_i\Delta v_i +(\partial_nv_n)^2\big>
\label{eqb14a}
\end{eqnarray}
and using a Transverse-Longitudinal (TL) decomposition\cite{pio}
with notation
\begin{eqnarray}
k_{00}\equiv n \, \, , \label{a1}
\end{eqnarray}
\begin{eqnarray}
h_{i0}=h_{0i} \equiv \partial_i k^L + \epsilon_{il}\partial_l
k^T\, \, , \label{a2}
\end{eqnarray}
\begin{eqnarray}
k_{ij}=k_{ji}\equiv (\eta_{ij}\Delta -\partial _i \partial
_j)k^{TT} +\partial _i \partial _j k^{LL}\nonumber
\\ +(\epsilon _{ik}\partial
_k\partial _j +\epsilon _{jk}\partial _k \partial _i )k^{TL}
\label{a3}
\end{eqnarray}
\begin{eqnarray}
v_0\equiv q \, \, , \label{a4}
\end{eqnarray}
\begin{eqnarray}
v_i \equiv \partial_i v^L + \epsilon_{il}\partial_l v^T\, \, ,
\label{a5}
\end{eqnarray}
where $\Delta \equiv \partial _i \partial _i$, eq. (\ref{eqb14a})
can be rewritten as follows
\begin{eqnarray}
{S^{(3)L}}_{0} =\kappa^2 \big<\dot{k}^L \Delta \dot{k}^L+\dot{k}^T
\Delta \dot{k}^T+\dot{v}^L \Delta \dot{v}^L+ \dot{v}^T \Delta
\dot{v}^T\nonumber \\+2\dot{v}^L \Delta \dot{k}^T-2\dot{v}^T
\Delta \dot{k}^L +(\Delta \dot{k}^{TT})^2 +(\Delta \dot{k}^{LL})^2
\nonumber \\+2(\Delta \dot{k}^{TL})^2+2(\dot{q})^2-2n\Delta
\dot{k}^L +2n\Delta \dot{v}^T\nonumber \\+2q\Delta
\dot{v}^L-2q\Delta \dot{k}^T+2\Delta k^{LL}\Delta
\dot{k}^L+2\Delta k^{TL}\Delta \dot{k}^T\nonumber \\+2\Delta
k^{LL}\Delta \dot{v}^T -2\Delta k^{TL}\Delta \dot{v}^L +q\Delta
q+n\Delta n\nonumber \\+(\Delta k^L)^2+2(\Delta k^T)^2 +2(\Delta
v^L)^2 +(\Delta v^T)^2\nonumber \\+2\Delta v^{T}\Delta k^{L}
+2q\Delta^2 k^{TL} +\Delta k^{TT}\Delta^2 k^{TT}\nonumber
\\+\Delta k^{TL}\Delta^2 k^{TL}\big>
\label{e14}
\end{eqnarray}

Primary Lagrangian constraints, joined to some links among
accelerations, can be obtained through an inspection on field
equations, which arise from (\ref{e14}). A ''Coulomb'' gauge is
defined by the constraints $\partial_ik_{i\mu}=0$, which can be
rewritten in terms of the TL-decomposition as follows (up to
harmonics)
\begin{equation}
k^L=k^{LL}=k^{TL}=0 \,\, , \label{e15}
\end{equation}
and preservation provides the next conditions for longitudinal
velocities and accelerations
\begin{equation}
\dot{k}^L=\dot{k}^{LL}=\dot{k}^{TL}=0 \,\, , \label{e16}
\end{equation}
\begin{equation}
\ddot{k}^L=\ddot{k}^{LL}=\ddot{k}^{TL}=0 \,\, . \label{e17}
\end{equation}
Equations (\ref{e15}) and (\ref{e16}) are six Lagrangian
constraints.

Field equations with the help of gauge constraints, give the
following  five (primary) constraints
\begin{equation}
n=0 \,\, , \label{e18}
\end{equation}
\begin{equation}
v^T=0 \,\, , \label{e19}
\end{equation}
\begin{equation}
\dot{v}^T=0 \,\, , \label{e20}
\end{equation}
\begin{equation}
\dot{k}^T-\dot{v}^L+q=0 \,\, , \label{e21}
\end{equation}
\begin{equation}
\Delta k^T-\Delta v^L+\dot{q}=0 \,\, , \label{e22}
\end{equation}
$n=\dot{n}=v^T=\dot{v}^T=0$\,\,$\dot{k}^T-\dot{v}^L+q=\Delta
k^T-\Delta v^L+\dot{q}=0$ and accelerations are related through
\begin{equation}
\ddot{v}^T=-\dot{n} \,\, , \label{e23}
\end{equation}
\begin{equation}
\ddot{k}^T+\ddot{v}^L=\dot{q} +2\Delta {k}^T\,\, , \label{e24}
\end{equation}
\begin{equation}
\ddot{k}^{TT}=\Delta k^{TT} \,\, , \label{e25}
\end{equation}
\begin{equation}
\ddot{q}=\Delta q \,\, . \label{e26}
\end{equation}

Systematic preservation of constraints (\ref{e18}) and (\ref{e22}) provide a new constraint
\begin{equation}
\dot{n}=0 \,\, , \label{e27}
\end{equation}
and accelerations
\begin{equation}
\ddot{n}=0 \,\, , \label{e28}
\end{equation}
\begin{equation}
\ddot{v}^T=0 \,\, , \label{e29}
\end{equation}
\begin{equation}
\ddot{k}^T=\Delta {k}^T\,\, , \label{e30}
\end{equation}
\begin{equation}
\ddot{v}^L=\Delta {v}^L \,\, . \label{e31}
\end{equation}

In short, there is a set of twelve constraints
\begin{equation}
n=\dot{n}=v^T=\dot{v}^T=k^L=\dot{k}^L=k^{LL}=\dot{k}^{LL}=k^{TL}=\dot{k}^{TL}=0 \,\, , \label{e32}
\end{equation}
\begin{equation}
\dot{k}^T-\dot{v}^L+q=0 \,\, , \label{e33}
\end{equation}
\begin{equation}
\Delta k^T-\Delta v^L+\dot{q}=0 \,\, , \label{e34}
\end{equation}
then, there are three degrees of freedom, and the constraint
system give rise to reduced action
\begin{eqnarray}
{S^{(3)L*}}_{0} =\kappa^2 \big<4\dot{k}^T \Delta
\dot{k}^T+4(\Delta k^T)^2 +4(\dot{q})^2\nonumber \\+4q\Delta q
+(\Delta \dot{k}^{TT})^2+\Delta k^{TT}\Delta^2 k^{TT}\big>
\label{e35}
\end{eqnarray}
Introducing notation
\begin{equation}
Q\equiv 2q \,\, , \label{e36}
\end{equation}
\begin{equation}
Q^T\equiv 2(-\Delta)^{\frac{1}2}k^T \,\, , \label{e37}
\end{equation}
\begin{equation}
Q^{TT}\equiv \Delta k^{TT} \,\, , \label{e38}
\end{equation}
the reduced action is rewritten as follows
\begin{eqnarray}
{S^{(3)L*}}_{0} =\kappa^2 \big<Q\Box{}Q-Q^T\Box{}Q^T+ Q^{TT}\Box{}Q^{TT}\big>\,\,,
\label{e35}
\end{eqnarray}
showing two unitary and one non-unitary modes, then the
Hamiltonian is not positive definite. This study could also have
considered from the point of view of the exchange amplitude
procedure, in which is considered the coupling to a (conserved)
energy-momentum tensor of some source, trough Lagrangian terms
$\kappa k_{\mu\nu}T^{\mu\nu}$ and $\chi v_{\mu}J^{\mu}$.

\vskip .2truein
\section{YM gravity with parity preserving massive term}

It can be possible to write down a massive version
which respect parity, for example
\begin{equation}
{S^{(3)}}_{m}={S^{(3)}}_0 - \frac{m^2
\kappa^2}{2}\big<{T^{\sigma}}_{\sigma
\nu}{{T^{\rho}}_{\rho}}^{\nu}-T^{\lambda\mu\nu}T_{\mu\lambda\nu}-\frac{1}{2}T^{\lambda\mu\nu}T_{\lambda\mu\nu}\big>\,
\, . \label{eqa3}
\end{equation}
In a general case, if we allow independent variations on metric and connection  two
types of field equations can be obtained. On one hand, variations
on metric give rise to the expression of the gravitacional
energy-momentum tensor, ${T_g}^{\alpha\beta}\equiv
\kappa^2\,tr[F^{\alpha\sigma}{F^\beta}_\sigma
-\frac{g^{\alpha\beta}}4 \, F^{\mu \nu}F_{\mu \nu}]$, in other
words
\begin{equation}
{T_g}^{\alpha\beta}=
-{T_t}^{\alpha\beta}-\kappa^2g^{\alpha\beta}\lambda^2\, \, ,
\label{eqa4}
\end{equation}
where ${T_t}^{\alpha\beta}\equiv
-m^2\kappa^2[3t^{\alpha\sigma}{t^\beta}_\sigma
+3t^{\sigma\alpha}{t_\sigma}^\beta
-t^{\alpha\sigma}{t_\sigma}^\beta-t^{\sigma\alpha}{t^\beta}_\sigma
- (t^{\alpha\beta}+t^{\beta\alpha}){t_\sigma}^\sigma
-\frac{5g^{\alpha\beta}}2 \, t^{\mu \nu}t_{\mu
\nu}+\frac{3g^{\alpha\beta}}2 \, t^{\mu \nu}t_{\nu
\mu}+\frac{g^{\alpha\beta}}2\,({t_\sigma}^\sigma)^2 ]$ is the
torsion contribution to the energy-momentum distribution and
$t^{\alpha\beta}\equiv\frac{\varepsilon^{\mu\nu\alpha}}2\,{T^{\beta}}_{\mu\nu}
$. This says, for example, that the quest of possible black hole
solutions must reveal a dependence on parameters  $m^2$ and
$\lambda^2$.

On the other hand, variations on connection provide the following
equations
\begin{eqnarray}
\frac 1{\sqrt{-g}}\,\,\partial _\alpha (\sqrt{-g}\,\,F^{\alpha
\lambda}) + [A_\alpha , F^{\alpha \lambda}] =J^\lambda \, \, ,
\label{eqa5}
\end{eqnarray}
where the current is $(J^\lambda)^{\nu}\,_\sigma =
m^2({\delta^\lambda}_\sigma {{K^\rho}_\rho}^\nu
-{\delta^\nu}_\sigma {{K^\rho}_\rho}^\lambda
+2{{K^\nu}_\sigma}^\lambda)$ and the contorsion
${K^\lambda}_{\mu\nu}\equiv \frac
1{2}({T^\lambda}_{\mu\nu}+{{T_\mu}^\lambda}_\nu +
{{T_\nu}^\lambda}_\mu)$. We can observe in (\ref{eqa5}) that
contorsion and metric appear as sources of gravity, where the
cosmological contribution is obviously hide in space-time metric.
In a weak torsion regime, equation (\ref{eqa5}) takes a familiar
shape, this means $\nabla_\alpha F^{\alpha \lambda} =J^\lambda $.

Now we explore the perturbation of the massive case given at
(\ref{eqa3}) and with the help of (\ref{eqb7}), the linearized
action is
\begin{eqnarray}
{S^{(3)L}}_m =\kappa^2 \big< k_{\mu\nu}\Box{}k^{\mu\nu}+
\partial_\mu k^{\mu\sigma}\partial_\nu {k^\nu}_\sigma
-2 \epsilon^{\sigma\alpha\beta}\partial_\alpha v_\beta
\partial_\nu {k^\nu}_\sigma\nonumber \\ - v_\mu \Box{}v^\mu +(\partial_\mu
v^\mu)^2-m^2(k_{\mu\nu}k^{\mu\nu}-k^2)\big> \,\, . \label{eqb15}
\end{eqnarray}

Using a TL-decomposition defined by (\ref{a1})-(\ref{a5}), we can
write  (\ref{eqb15}) in the way
\begin{eqnarray}
{S^{(3)L}}_m =\kappa^2 \big<\dot{k}^L \Delta \dot{k}^L+\dot{k}^T
\Delta \dot{k}^T+\dot{v}^L \Delta \dot{v}^L+ \dot{v}^T \Delta
\dot{v}^T+2\dot{v}^L \Delta \dot{k}^T-2\dot{v}^T \Delta \dot{k}^L
\nonumber \\
+(\Delta \dot{k}^{TT})^2 +(\Delta \dot{k}^{LL})^2 +2(\Delta
\dot{k}^{TL})^2+2(\dot{q})^2-2n\Delta \dot{k}^L +2n\Delta
\dot{v}^T\nonumber
\\+2q\Delta \dot{v}^L-2q\Delta
\dot{k}^T+2\Delta k^{LL}\Delta \dot{k}^L+2\Delta k^{TL}\Delta
\dot{k}^T+2\Delta k^{LL}\Delta \dot{v}^T \nonumber \\-2\Delta
k^{TL}\Delta \dot{v}^L +q\Delta q+n\Delta n+(\Delta
k^L)^2+2(\Delta k^T)^2 +2(\Delta v^L)^2\nonumber \\ +(\Delta
v^T)^2+2\Delta v^{T}\Delta k^{L} +2q\Delta^2 k^{TL} +\Delta
k^{TT}\Delta^2 k^{TT}+\Delta k^{TL}\Delta^2 k^{TL}\nonumber
\\+m^2[-2k^L\Delta k^L-2k^T\Delta k^T-2(\Delta  k^{TL})^2 -2n(\Delta  k^{TT}+\Delta  k^{LL})\nonumber
\\+2\Delta  k^{TT}\Delta  k^{LL}]\big>\,\,. \label{eqba1}
\end{eqnarray}
Here, there is no gauge freedom (as it shall be confirmed in next
section) and field equations provide primary constraints and some
accelerations. The preservation procedure gives rise to
expressions for all accelerations
\begin{equation}
\ddot{n}=\Delta \dot{k}^L \,\, , \label{eqba2}
\end{equation}
\begin{equation}
\ddot{k}^L=-\Delta \dot{k}^{TT}+\dot{n} \,\, , \label{eqba3}
\end{equation}
\begin{equation}
\ddot{k}^T=\Delta \dot{k}^{TL}\,\, , \label{eqba4}
\end{equation}
\begin{equation}
\ddot{k}^{LL}= \dot{k}^L+ \dot{v}^T+m^2k^{TT}-m^2\Delta^{-1} n
\,\, , \label{eqba5}
\end{equation}
\begin{equation}
\ddot{k}^{TL}=\frac{1}2 (\dot{k}^T- \dot{v}^L+\Delta
k^{TL}+q-2m^2k^{TL}) \,\, , \label{eqba6}
\end{equation}
\begin{equation}
\ddot{k}^{TT}=\Delta k^{TT}+m^2k^{LL}-m^2\Delta^{-1} n\,\, ,
\label{eqba7}
\end{equation}
\begin{equation}
\ddot{q}=\frac{1}2 (\Delta\dot{v}^L-\Delta \dot{k}^T+\Delta^2
k^{TL}+\Delta q) \,\, , \label{eqba8}
\end{equation}
\begin{equation}
\ddot{v}^L=-\dot{q}+2\Delta v^L\,\, , \label{eqba9}
\end{equation}
\begin{equation}
\ddot{v}^T=-\Delta \dot{k}^{TT}-\Delta \dot{k}^{LL}+\Delta
k^L+\Delta v^T \,\, , \label{eqba10}
\end{equation}
and a set of eight constraints
\begin{equation}
\dot{v}^T-\dot{k}^L+n -m^2(k^{TT}+k^{LL})=0\,\, , \label{c1}
\end{equation}
\begin{equation}
\Delta \dot{k}^{LL}-\Delta k^L-\Delta v^T +m^2k^L=0\,\, ,
\label{c2}
\end{equation}
\begin{equation}
\Delta \dot{k}^{TL}-\dot{q}-\Delta k^T+\Delta v^L +m^2k^T=0 \,\, ,
\label{c3}
\end{equation}
\begin{equation}
\Delta \dot{k}^{LL}-\Delta k^L-\Delta v^T
+m^2(\dot{k}^{TT}+\dot{k}^{LL})=0\,\, , \label{c4}
\end{equation}
\begin{equation}
\dot{k}^L+\Delta k^{TT}-n =0 \,\, , \label{c5}
\end{equation}
\begin{equation}
\dot{k}^T-\Delta k^{TL}=0\,\, , \label{c6}
\end{equation}
\begin{equation}
\dot{v}^T+\Delta k^{TT}+m^2(k^{TT}+k^{LL})-2m^2\Delta^{-1} n=0
\,\, , \label{c7}
\end{equation}
\begin{equation}
\dot{n}-\Delta k^{L}=0 \,\, , \label{c8}
\end{equation}
which says that this massive theory propagates five degrees of
freedom. In order to explore the physical content, we can take a
short path to this purpose and it means to start with a typical
transverse-traceless (Tt) decomposition instead the
TL-decomposition one. Notation for the Tt-decomposition of fields
is
\begin{equation}
k_{\mu\nu}={k^{Tt}}_{\mu\nu}+\hat{\partial}_\mu {\theta^{T}}_\nu
+\hat{\partial}_\nu
{\theta^{T}}_\mu+\hat{\partial}_\mu\hat{\partial}_\nu \psi
+\eta_{\mu\nu}\phi \,\, , \label{eqb16}
\end{equation}
\begin{equation}
v_\mu={v^{T}}_\mu+\hat{\partial}_\mu v \,\, , \label{eqb17}
\end{equation}
with the subsidiary conditions
\begin{equation}
{k^{Tt\mu}}_\mu=0\,\,\,\,,\,\,\,\,\partial^\mu{k^{Tt}}_{\mu\nu}=0\,\,\,\,,\,\,\,\,\partial^\mu
{\theta^{T}}_\mu =0\,\,\,\,,\,\,\,\,\partial^\mu {v^{T}}_\mu =0
\,\, . \label{eqb18}
\end{equation}
Action (\ref{eqb15}) is
\begin{eqnarray}
{S^{(3)L}}_m =\kappa^2 \big<
{k^{Tt}}_{\mu\nu}(\Box{}-m^2){k^{Tt}}^{\mu\nu}-{\theta^{T}}_\mu(\Box{}-2m^2){\theta^{T}}_\mu
-2 \epsilon^{\sigma\alpha\beta}\partial_\alpha {v^{T}}_\beta
\Box{}^{\frac{1}2}{\theta^{T}}_\sigma\nonumber \\ - {v^{T}}_\mu
\Box{}{v^{T}}^\mu
+2v\Box{}v+2\phi\Box{}\phi+4m^2\psi\phi+6m^2\phi^2\big> \,\, .
\label{d1}
\end{eqnarray}
A new transverse variable, ${a^T}_\mu$ is introduced through
\begin{equation}
{\theta^{T}}_\mu \equiv
{\epsilon_\mu}^{\alpha\beta}\hat{\partial}_\alpha
{a^{T}}_\beta\,\, , \label{d2}
\end{equation}
and the action (\ref{d1}) is rewritten as
\begin{eqnarray}
{S^{(3)L}}_m =\kappa^2 \big<
{k^{Tt}}_{\mu\nu}(\Box{}-m^2){k^{Tt}}^{\mu\nu}-{a^{T}}_\mu(\Box{}-2m^2){a^{T}}_\mu
-2 {a^{T}}_\mu\Box{}{v^{T}}^\mu \nonumber \\ - {v^{T}}_\mu
\Box{}{v^{T}}^\mu
+2v\Box{}v+2\phi\Box{}\phi+4m^2\psi\phi+6m^2\phi^2\big> \,\, .
\label{d3}
\end{eqnarray}
The field equations are
\begin{equation}
(\Box{}-m^2){k^{Tt}}_{\mu\nu}=0 \,\, , \label{d4}
\end{equation}
\begin{equation}
\Box{} {v^{T}}_\mu=0 \,\, , \label{d5}
\end{equation}
\begin{equation}
\Box{}v=0\,\, , \label{d6}
\end{equation}
\begin{equation}
{a^{T}}_\mu=0 \,\, , \label{d7}
\end{equation}
\begin{equation}
 \psi=\phi=0 \,\, , \label{d8}
\end{equation}
and reduced action is
\begin{eqnarray}
{S^{(3)L*}}_m =\kappa^2 \big<
{k^{Tt}}_{\mu\nu}(\Box{}-m^2){k^{Tt}}^{\mu\nu}+2v\Box{}v-
{v^{T}}_\mu \Box{}{v^{T}}^\mu \big> \,\, , \label{d3}
\end{eqnarray}
saying that the contorsion propagates two massive helicities $\pm
2$, one massless spin-$0$ and two massless ghost vectors. Then,
there is not positive definite Hamiltonian. This observation can
be confirmed in the next section when we shall write down the
Hamiltonian density and a wrong sign appears in the kinetic part
corresponding to the canonical momentum of $v_i$ (see eq.
(\ref{eqb33})).

\vskip .2truein
\section{Gauge transformations}

The quadratical Lagrangian density dependent in torsion and
presented in (\ref{eqa3}), has been constructed without free
parameters, with the exception of $m^2$, of course. It has a
particular shape which only gives mass to the spin 2 component of
the contorsion, as we see in the perturbative regime. Let us
comment about de non existence of any possible ''residual'' gauge
invariance of the model. The answer is that the model lost its
gauge invariance and it can be shown performing the study of
symmetries through computation of the gauge generator chains. For
this purpose, a $2+1$ decomposition of (\ref{eqb15}) is performed,
this means
\begin{eqnarray}
{S^{(3)L}}_{m} =\kappa^2
\big<[-\dot{k}_{0i}+2\partial_ik_{00}-2\partial_nk_{ni}-2\epsilon_{in}\dot{v}_n
+2\epsilon_{in}\partial_n v_0
]\dot{k}_{0i}+\dot{k}_{ij}\dot{k}_{ij}\nonumber \\
+[2\epsilon_{nj}\partial_nk_{00}+2\epsilon_{nj}\partial_mk_{nm}-\dot{v}_j-2\partial_jv_0]\dot{v}_j+2(\dot{v}_0)^2
+k_{00}\Delta k_{00}\nonumber \\-2k_{0i}\Delta k_{0i}+k_{ij}\Delta
k_{ij}-(\partial_ik_{i0})^2+\partial_nk_{ni}\partial_mk_{mi}
-2\epsilon_{ij}\partial_i v_j
\partial_nk_{n0}\nonumber \\-2\epsilon_{lm}\partial_m v_0
\partial_nk_{nl}+v_0\Delta v_0-v_i\Delta v_i +(\partial_nv_n)^2\nonumber \\
+m^2[2k_{0i}k_{0i}-k_{ij}k_{ij}-2k_{00}k_{ii}+(k_{ii})^2]\big>\,\,,
\label{eqb22}
\end{eqnarray}
where $\epsilon_{ij}\equiv {\epsilon^0}_{ij}$ and $\Delta\equiv
\partial_i\partial_i$.

Next, the  momenta are
\begin{equation}
\Pi\equiv \frac{\partial\mathcal{L}}{\partial\dot{k}_{00}}=0\,\, ,
\label{eqb23}
\end{equation}
\begin{equation}
\Pi^i\equiv
\frac{\partial\mathcal{L}}{\partial\dot{k}_{0i}}=-2\dot{k}_{0i}-2\epsilon_{in}
\dot{v}_n+2\partial_ik_{i0}-2\partial_nk_{ni}+2\epsilon_{in}
\partial_nv_0\,\, , \label{eqb24}
\end{equation}
\begin{equation}
\Pi^{ij}\equiv
\frac{\partial\mathcal{L}}{\partial\dot{k}_{ij}}=2\dot{k}_{ij}\,\,
, \label{eqb25}
\end{equation}
\begin{equation}
P\equiv
\frac{\partial\mathcal{L}}{\partial\dot{v}_0}=4\dot{v}_0\,\, ,
\label{eqb26}
\end{equation}
\begin{equation}
P^j\equiv
\frac{\partial\mathcal{L}}{\partial\dot{v}_j}=-2\epsilon_{nj}\dot{k}_{0n}-2\dot{v}_j+2\epsilon_{nj}\partial_nk_{00}
+2\epsilon_{nj}\partial_mk_{mn} -2\partial_jv_0\,\, ,
\label{eqb27}
\end{equation}
and we establish the following commutation rules
\begin{equation}
\{k_{00}(x),\Pi (y)\}=\{v_0(x),P(y)\}=\delta^2(x-y)\,\, ,
\label{eqb28}
\end{equation}
\begin{equation}
\{k_{0i}(x),\Pi^j (y)\}=\{v_i(x),P^j(y)\}={\delta^j}_i
\delta^2(x-y)\,\, , \label{eqb29}
\end{equation}
\begin{equation}
\{k_{ij}(x),\Pi^{nm}(y)\}=\frac{1}2({\delta^n}_i{\delta^m}_j+{\delta^m}_i{\delta^n}_j)
\delta^2(x-y)\,\,. \label{eqb30}
\end{equation}

It can be noted that (\ref{eqb23}) is a primary constraint that we
name
\begin{equation}
G^{(K)}\equiv\Pi\,\, , \label{eqb31}
\end{equation}
where $K$ means the initial index corresponding to a possible
gauge generator chain, provided by the algorithm developed in
reference\cite{c}. Moreover, manipulating (\ref{eqb25}) and
(\ref{eqb27}), other primary constraints appear
\begin{equation}
G^{(K)}_i\equiv \partial_n
k_{ni}-\epsilon_{in}\partial_nv_0-\frac{\epsilon_{in}}{4}P^n+\frac{1}{4}\Pi^i\,\,
, \label{eqb32}
\end{equation}
and we observe that $G^{(K)}$ and $G^{(K)}_i$ are first class.

The preservation of constraints requires to obtain the Hamiltonian
of the model. First of all, the Hamiltonian density can be written
as
$\mathcal{H}_0=\Pi^i\dot{h}_{0i}+\Pi^{ij}\dot{h}_{ij}+P\dot{v}_0+P^i\dot{v}_i-\mathcal{L}$,
in other words
\begin{eqnarray}
\mathcal{H}_0 =\frac{\Pi^{ij}\Pi^{ij}}{4}+ \frac{P^2}{8}
-\frac{P^iP^i}{4}+\epsilon_{nj}\partial_mk_{nm}P^j+v_0[\partial_iP^i+4\epsilon_{ml}\partial_m\partial_nk_{nl}]\nonumber
\\+k_{00}[2\partial_m\partial_nk_{mn}-\epsilon_{nm}\partial_nP^m+2m^2k_{ii}]
+2k_{0i}\Delta k_{0i}-k_{ij}\Delta k_{ij}\nonumber
\\+(\partial_ik_{i0})^2-2\partial_nk_{ni}\partial_mk_{mi}
+2\epsilon_{ij}\partial_i v_j
\partial_nk_{n0}+v_i\Delta v_i -(\partial_nv_n)^2\nonumber \\
-m^2[2k_{0i}k_{0i}-k_{ij}k_{ij}+(k_{ii})^2]\,\,. \label{eqb33}
\end{eqnarray}

Then, the Hamiltonian is $H_0=\int dy^2 \mathcal{H}_0(y)\equiv
\big<\mathcal{H}_0\big>_y$ and the preservation of $G^{(K)}$,
defined in (\ref{eqb31}) is
\begin{equation}
\{G^{(K)}(x),H_0\}=-2\partial_m\partial_nk_{mn}(x)+\epsilon_{nm}\partial_nP^m(x)-2m^2k_{ii}(x)\,\,
. \label{eqb34}
\end{equation}
The possible generators chain is given by the rule:
''$G^{(K-1)}+\{G^{(K)}(x),H_0\}=${\it combination of primary
constraints}'', then
\begin{eqnarray}
G^{(K-1)}(x)=2\partial_m\partial_nk_{mn}(x)-\epsilon_{nm}\partial_nP^m(x)+2m^2k_{ii}(x)
\nonumber \\+\big<a(x,y)G^{(K)}(y)+b^i(x,y)G^{(K)}_i(y)\big>_y\,\,
. \label{eqb35}
\end{eqnarray}

The preservation of $G^{(K)}_i$, defined in (\ref{eqb32}), is
\begin{eqnarray}
\{G^{(K)}_i(x),H_0\}=\frac{\partial_n\Pi^{ni}(x)}{2}-\frac{\epsilon_{in}}{4}\partial_nP(x)+\frac{\epsilon_{in}}{2}\Delta
v_n(x) +\frac{\epsilon_{in}}{2}\partial_n\partial_mv_m(x)\nonumber
\\+\frac{\epsilon_{nm}}{2}\partial_i\partial_nv_m(x)-(\Delta
-m^2)k_{0i}(x) \,\, , \label{eqb36}
\end{eqnarray}
then
\begin{eqnarray}
G^{(K-1)}_i(x)=-\frac{\partial_n\Pi^{ni}(x)}{2}+\frac{\epsilon_{in}}{4}\partial_nP(x)-\frac{\epsilon_{in}}{2}\Delta
v_n(x) -\frac{\epsilon_{in}}{2}\partial_n\partial_mv_m(x)\nonumber
\\-\frac{\epsilon_{nm}}{2}\partial_i\partial_nv_m(x)+(\Delta
-m^2)k_{0i}(x)\nonumber
\\
+\big<a^i(x,y)G^{(K)}(y)+{b^i}_j(x,y)G^{(K)}_j(y)\big>_y\,\, .
\label{eqb37}
\end{eqnarray}

The undefined objects $a(x,y)$, $b^i(x,y)$, $a^i(x,y)$ and
${b^i}_j(x,y)$ in  expressions (\ref{eqb35}) and (\ref{eqb37}),
are functions or distributions. If it is possible, they can be
fixed in a way that the preservation of  $G^{(K-1)}(x)$ and
$G^{(K-1)}_i(x)$ would be combinations of primary constraints.
With this, the generator chains could be interrupted and we simply
take $K=1$. Of course, the order $K-1=0$ generators must be first
class, as every one. Next, we can see that all these statements
depend on the massive or non-massive character of the theory.

Taking a chain with $K=1$, the candidates to generators of gauge
transformation are (\ref{eqb31}), (\ref{eqb32}), (\ref{eqb35}) and
(\ref{eqb37}). But, the only non null commutators are
\begin{eqnarray}
\{G^{(1)}_i(x),G^{(0)}_j(y)\}=\frac{m^2}{4}\eta_{ij}\delta^2(x-y)
\,\, , \label{eqb38}
\end{eqnarray}
\begin{eqnarray}
\{G^{(0)}(x),G^{(0)}_i(y)\}=m^2\big(\partial_i\delta^2(x-y)+\frac{b^i(x,y)}{4}\big)
\,\, , \label{eqb39}
\end{eqnarray}
saying that the system of ''generators'' is not first class.
Moreover, the unsuccessful conditions (in the $m^2\neq 0$ case) to
interrupt the chains, are
\begin{eqnarray}
\{G^{(0)}(x),H_0\}=m^2(\Pi^{nn}(x)-2\partial_nk_{0n}(x)) \,\, ,
\label{eqb40}
\end{eqnarray}
\begin{eqnarray}
\{G^{(0)}_i(x),H_0\}=m^2(\partial_nk_{in}(x)+\partial_ik_{00}(x)-\partial_ik_{nn}(x))
\,\, , \label{eqb41}
\end{eqnarray}
where we have fixed
\begin{eqnarray}
a(x,y)=0 \,\, , \label{eqb42}
\end{eqnarray}
\begin{eqnarray}
b^i(x,y)=-2\partial^i\delta^2(x-y) \,\, , \label{eqb43}
\end{eqnarray}
\begin{eqnarray}
a^i(x,y)=0 \,\, , \label{eqb44}
\end{eqnarray}
\begin{eqnarray}
{b^i}_j(x,y)=0 \,\, . \label{eqb45}
\end{eqnarray}

All this indicates that in the case where $m^2\neq 0$ there is not
a first class consistent chain of generators and, then there is no
gauge symmetry.

However, if we revisit the case $m^2=0$, conditions (\ref{eqb40})
and (\ref{eqb41}) are zero and the chains are interrupted. Now,
the generators $G^{(1)}$, $G^{(1)}_i$, $G^{(0)}$ and $G^{(0)}_i$
are first class. Using (\ref{eqb42})-(\ref{eqb43}), the generators
are rewritten again
\begin{equation}
G^{(1)}\equiv\Pi\,\, , \label{eqb46}
\end{equation}
\begin{equation}
G^{(1)}_i\equiv \partial_n
k_{ni}-\epsilon_{in}\partial_nv_0-\frac{\epsilon_{in}}{4}P^n+\frac{1}{4}\Pi^i\,\,
, \label{eqb47}
\end{equation}
\begin{eqnarray}
G^{(0)}=-\frac{\epsilon_{nm}}{2}\partial_nP^m-\frac{\partial_n\Pi^n}{2}
\,\, , \label{eqb48}
\end{eqnarray}
\begin{eqnarray}
G^{(0)}_i=-\frac{\partial_n\Pi^{ni}}{2}+\frac{\epsilon_{in}}{4}\partial_nP-\frac{\epsilon_{in}}{2}\Delta
v_n
-\frac{\epsilon_{in}}{2}\partial_n\partial_mv_m-\frac{\epsilon_{nm}}{2}\partial_i\partial_nv_m+\Delta
k_{0i}\,\, . \label{eqb49}
\end{eqnarray}

Introducing the parameters $\varepsilon (x)$ and $\varepsilon^i
(x)$, a combination of (\ref{eqb46})-(\ref{eqb49}) is taken into
account in the way that the gauge generator is
\begin{eqnarray}
G(\dot{\varepsilon} , \dot{\varepsilon}^i , \varepsilon ,
\varepsilon^i)=\big<
\dot{\varepsilon}(x)G^{(1)}(x)+\dot{\varepsilon}^i(x)G^{(1)}_i(x)+\varepsilon(x)
G^{(0)}(x) +\varepsilon^i(x)G^{(0)}_i(x)\big>\,\, , \label{eqb50}
\end{eqnarray}
and with this, for example the field transformation rules (this
means,  $\delta (...)= \{(...), G \}$) are written as
\begin{eqnarray}
\delta k_{00}= \dot{\varepsilon}\,\, , \label{eqb51}
\end{eqnarray}
\begin{eqnarray}
\delta k_{0i}=
\frac{\dot{\varepsilon}^i}{4}+\frac{\partial_i\varepsilon}{2}\,\,
, \label{eqb52}
\end{eqnarray}
\begin{eqnarray}
\delta k_{ij}=
\frac{1}{4}(\partial_i\varepsilon_j+\partial_j\varepsilon_i)\,\, ,
\label{eqb53}
\end{eqnarray}
\begin{eqnarray}
\delta v_0= \frac{\epsilon_{nm}}{4}\partial_n\varepsilon_m\,\, ,
\label{eqb54}
\end{eqnarray}
\begin{eqnarray}
\delta v_i= \frac{\epsilon_{in}}{4}\dot{\varepsilon}_n
-\frac{\epsilon_{in}}{2}\partial_n\varepsilon\,\, , \label{eqb55}
\end{eqnarray}
and, redefining parameters as follows: $\varepsilon \equiv 2\xi_0$
and $\varepsilon^i= 4 \xi^i$, it is very easy to see that these
rules match with (\ref{eqb9}) and (\ref{eqb10}), as we expected.

\vskip .2truein
\section{YM-extended formulation}

Here we review a possible  quadratical term family which allows to
eliminate non-unitary propagations in the contorsion (torsion)
perturbative regime in $2+1$ dimension. The most general shape of
a Lagrangian counter terms set is
\begin{eqnarray}
{S^{(3)}}_0=\kappa^{2} \big<-\frac 14 \,
{(F^{\mu\nu})^\sigma}_\rho {(F_{\mu\nu})^\rho}_\sigma +
a_1{(F_{\mu\nu})^\sigma}_\rho ({F^\mu}_\sigma )^{\nu\rho}+ a_2
{(F_{\mu\nu})^\sigma}_\rho ({F_\sigma}^\rho )^{\mu\nu}\nonumber
\\ + a_3{(F_{\mu\sigma})^\sigma}_\nu
({F^\mu}_\rho )^{\rho\nu}+a_4{(F_{\mu\sigma})^\sigma}_\nu
({F^\nu}_\rho )^{\rho\mu}+a_5((F_{\mu\nu})^{\mu\nu})^2\big> \,\,
,\,\, \label{eqc1}
\end{eqnarray}
where $a_1$, $a_2$, $a_3$, $a_4$ and $a_5$ are real parameters.

A naive try to reach unitarity consists to perform a direct
matching between the perturbative action coming from (\ref{eqc1})
and the linearized Hilbert-Einstein one, given by
\begin{eqnarray}
{S_{HE}}^L =-2\kappa^2 \big<
h_{\mu\nu}{G_L}^{\mu\nu}\big>=\kappa^2 \big<
h_{\mu\nu}\Box{}h^{\mu\nu}+ 2\partial_\mu
h^{\mu\sigma}\partial_\nu {h^\nu}_\sigma +2h\partial_\mu
\partial_\nu h^{\mu\nu}-h\Box{}h\big>\,\, ,\,\, \label{eqc2}
\end{eqnarray}
where $h_{\mu\nu}$ is the metric perturbation and ${G_L}^{\mu\nu}$
is the linearized Einstein's tensor. Then, under perturbations of
the contorsion (torsion), one can use again eq. (\ref{eqb7}), this
time in (\ref{eqc1}). Next, making comparison with (\ref{eqc2}), a
two free parameter system can be obtained (i.e., $a_3\equiv
\alpha$ and $a_5\equiv \beta$) and possible unitary theories which
propagates massless spin 2 in 2+1 dimension, are
\begin{eqnarray}
{S^{(3)}}_{(\alpha,\beta)}=\kappa^{2} \big<-\frac 14 \,
{(F^{\mu\nu})^\sigma}_\rho {(F_{\mu\nu})^\rho}_\sigma
-(1+\alpha){(F_{\mu\nu})^\sigma}_\rho ({F^\mu}_\sigma
)^{\nu\rho}\nonumber
\\ + (\frac58 +\frac{\alpha}2 +\beta) {(F_{\mu\nu})^\sigma}_\rho
({F_\sigma}^\rho )^{\mu\nu} + \alpha {(F_{\mu\sigma})^\sigma}_\nu
({F^\mu}_\rho )^{\rho\nu}\nonumber
\\-(\frac12 +\alpha +4\beta){(F_{\mu\sigma})^\sigma}_\nu
({F^\nu}_\rho )^{\rho\mu}+\beta((F_{\mu\nu})^{\mu\nu})^2\big> \,\,
,\,\, \label{eqc3}
\end{eqnarray}
and they are labeled with parameters  $\alpha$ and $\beta$. There
are two possible massives cases. On one hand, can be considered
the topological massive model (\ref{eqa2}), which is sensitive
under parity. On the other hand, there is a ''Fierz-Pauli'' model
(\ref{eqa3}). Our main pourpose in this section is to study the
classical consistence of field equations, focusing the attention
at the massless and topological massive cases.

In the massless theory with cosmological constant, $\lambda$ in
$2+1$ dimension, we introduce a cosmological term as follows
\begin{eqnarray}
{S^{(3)}}_{(\alpha,\beta,
\lambda)}={S^{(3)}}_{(\alpha,\beta)}+\kappa^{2}
\big<q(\alpha,\beta)\lambda^2\big> \,\, ,\,\, \label{eqc4}
\end{eqnarray}
where $q(\alpha,\beta)$ is a (unknown) real function of family's
parameters. Next, in order to consider classical consistence at
the torsionless regime, we take into account some auxiliary fields
(Lagrange multipliers), $b_{\mu\nu}$ and the action with torsion
constraints is given by
\begin{eqnarray}
{S'^{(3)}}_{(\alpha,\beta,
\lambda)}={S^{(3)}}_{(\alpha,\beta)}+\kappa^{2}
\big<q(\alpha,\beta)\lambda^2\big>+\kappa^2\big<b_{\alpha\beta}\,\varepsilon^{\beta\lambda\sigma}
{(A_\lambda)^\alpha}_\sigma \big>\,\, ,\,\, \label{eqc5}
\end{eqnarray}
where arbitrary variations on fields $b_{\mu\nu}$, obviously
provide the condition ${T^\alpha}_{\mu\nu}=0$. Then, the field
equation coming from variations of connection is
\begin{eqnarray}
\nabla_\mu {(\mathcal{F}^{\mu\nu})^\sigma}_\rho+
b_{\rho\mu}\,\varepsilon^{\mu\nu\sigma}=0\,\, ,\,\, \label{eqc5a}
\end{eqnarray}
where $\mathcal{F}_{\mu\nu}$ is defined in terms of the Yang-Mills
curvature, $F_{\mu\nu}$ in the way
\begin{eqnarray}
{(\mathcal{F}^{\mu\nu})^\sigma}_\rho\equiv
{(F^{\mu\nu})^\sigma}_\rho +2(1+\alpha) [({F^\mu}_\rho
)^{\nu\sigma}-({F^\nu}_\rho )^{\mu\sigma}] + (\frac54 +\alpha
+2\beta)[({F_\rho}^\sigma )^{\nu\mu} -({F_\rho}^\sigma )^{\mu\nu}]
\nonumber
\\+2 \alpha [({F^\nu}_\lambda
)^{\lambda\sigma} {\delta^\mu}_\rho-({F^\mu}_\lambda
)^{\lambda\sigma} {\delta^\nu}_\rho ]+(1 +2\alpha
+8\beta)[({F^\sigma}_\lambda )^{\lambda\mu}
{\delta^\nu}_\rho-({F^\sigma}_\lambda )^{\lambda\nu}
{\delta^\mu}_\rho ]\nonumber
\\+2\beta(F_{\lambda\kappa})^{\lambda\kappa}(g^{\mu\sigma}{\delta^\nu}_\rho -g^{\nu\sigma}{\delta^\mu}_\rho
)\,\, ,\,\,\nonumber
\\ \label{eqc5b}
\end{eqnarray}
and now, we can match the YM curvature with the
Riemann-Christoffel one (i. e.,
$(F_{\mu\nu})_{\alpha\beta}=R_{\alpha\beta\nu\mu}$), which
satisfies the well known algebraic properties and Bianchi
identities recalling as follows
\begin{eqnarray}
Symmetry:\,\,\,\,R_{\alpha\beta\nu\mu}=R_{\nu\mu\alpha\beta}\,\,
,\,\, \label{eqc6a}
\end{eqnarray}
\begin{eqnarray}
Antisymmetry:\,\,\,\,R_{\alpha\beta\nu\mu}=-R_{\beta\alpha\nu\mu}=R_{\beta\alpha\mu\nu}=-R_{\alpha\beta\mu\nu}\,\,
,\,\, \label{eqc6b}
\end{eqnarray}
\begin{eqnarray}
Cyclicity:\,\,\,\,R_{\alpha\beta\nu\mu}+R_{\alpha\mu\beta\nu}+R_{\alpha\nu\mu\beta}=0\,\,
,\,\, \label{eqc6c}
\end{eqnarray}
\begin{eqnarray}
Bianchi\,\,identities:\,\,\,\,\nabla_\sigma
R_{\alpha\beta\nu\mu}+\nabla_\mu
R_{\alpha\beta\sigma\nu}+\nabla_\nu R_{\alpha\beta\mu\sigma}=0\,\,
.\,\, \label{eqc6d}
\end{eqnarray}

In $2+1$ dimension, the curvature tensor can be written in terms
of Ricci's tensor ($R_{\mu\sigma}\equiv
{R^\lambda}_{\mu\lambda\sigma}$) and its trace ($R\equiv
{R^\lambda}_\lambda$) in the way
$R_{\lambda\mu\nu\sigma}=g_{\lambda\nu}R_{\mu
\sigma}-g_{\lambda\sigma}R_{\mu\nu}-g_{\mu
\nu}R_{\lambda\sigma}+g_{\mu\sigma}R_{\lambda\nu}-\frac{R}{2}\,
(g_{\lambda\nu}g_{\mu \sigma}-g_{\lambda\sigma}g_{\mu\nu})$. So,
the object defined in (\ref{eqc5b}) takes the shape
\begin{eqnarray}
(\mathcal{F}_{\sigma\nu})_{\lambda\mu}=(\frac32 +4\beta
)R_{\lambda\mu\nu\sigma}+(1+8\beta)(g_{\mu\nu}R_{\lambda
\sigma}-g_{\mu\sigma}R_{\lambda\nu})\nonumber \\+\,2\beta R
(g_{\lambda\nu}g_{\mu \sigma}-g_{\lambda\sigma}g_{\mu\nu})\,\,
,\,\, \label{eqc7}
\end{eqnarray}
which do not depend on parameter $\alpha$. Moreover, if $\beta$ is
fixed as
\begin{eqnarray}
\beta= -\frac18\,\, ,\,\, \label{eqc7a}
\end{eqnarray}
then, relation (\ref{eqc7}) leads to
\begin{eqnarray}
(\mathcal{F}_{\sigma\nu})_{\lambda\mu}\mid_{\beta=
-\frac18}=R_{\lambda\mu\nu\sigma}-\frac{R}4 (g_{\lambda\nu}g_{\mu
\sigma}-g_{\lambda\sigma}g_{\mu\nu})\,\, ,\,\, \label{eqc7b}
\end{eqnarray}
and this one satisfies all symmetry properties of a curvature,
showing in relations (\ref{eqc6a})-(\ref{eqc6c}) with the
exception of the Bianchi identities, (\ref{eqc6d}). It can be
noted that the trace of (\ref{eqc7b}), this means
${(\mathcal{F}_{\sigma\lambda})^\lambda}_\mu$ is the Einstein's
tensor.

Next, some discussion on the critical value (\ref{eqc7a}) shall be
performed when the connection's field equation is taking into
account. With the help of symmetry properties, Bianchi's
identities, and relationship between Riemann-Christoffel and Ricci
tensor, the field equation (\ref{eqc5a}) can be rewritten as
follows
\begin{eqnarray}
(\frac12-4\beta)\nabla_\rho
R_{\nu\sigma}-(\frac32+4\beta)\nabla_\sigma
R_{\nu\rho}+(\frac12+2\beta)g_{\nu\rho}\partial_\sigma R +2\beta
g_{\nu\sigma}\partial_\rho R\nonumber
\\+\,
b_{\rho\mu}\,{\varepsilon^\mu}_{\nu\sigma}=0\,\, ,\,\,
\label{eqc8}
\end{eqnarray}
and with some algebraic computation, it can be shown (for all
$\beta$) the next symmetry property
\begin{eqnarray}
b_{\nu\mu}= b_{\mu\nu}\,\, ,\,\, \label{eqc8a}
\end{eqnarray}
and
\begin{eqnarray}
(\beta-\frac58)b_{\mu\nu}= 0\,\, ,\,\, \label{eqc8b}
\end{eqnarray}
\begin{eqnarray}
(\beta+\frac18)\partial_\mu R=0\,\, .\,\, \label{eqc8c}
\end{eqnarray}
Consistence condition (\ref{eqc8b}) stablishes that the work out
of Lagrange multipliers depends on the following restriction
\begin{eqnarray}
\beta\neq \frac58\,\, ,\,\, \label{eqc9}
\end{eqnarray}
then, $b_{\mu\nu}=0$. Condition (\ref{eqc9}) induces a wide set of
possible vacuum's solutions, including non-Einstenian ones beside
(A)dS, because eq. (\ref{eqc8c}) becomes an identity when it is
evaluated on the critical $\beta$ given by (\ref{eqc7a}). This
fact is confirmed when  $\beta=-\frac18$ is introduced in eq.
(\ref{eqc8}), in other words
\begin{eqnarray}
\nabla_\rho \mathcal{R}_{\nu\sigma}-\nabla_\sigma
\mathcal{R}_{\nu\rho}=0\,\, ,\,\, \label{eqc9a}
\end{eqnarray}
where notation means
\begin{eqnarray}
\mathcal{R}_{\mu\nu}\equiv R_{\mu\nu}- \frac{g_{\mu\nu}}4\,R\,\,
.\,\, \label{eqc9b}
\end{eqnarray}
It can be observed that equation (\ref{eqc9a}) looks like eq.
(\ref{g1a}), but here, as one can expect the trace
$\sigma-\lambda$ of (\ref{eqc9a}) is an identity.

In order to conclude the comments on the massless theory, next we
consider the field equation which comes from variations on metric
of the action (\ref{eqc5}) and it can be written in terms of
Ricci's tensor and Ricci's scalar as follows
\begin{eqnarray}
(\frac32-\alpha+12\beta)R_{\sigma\mu}{R^\sigma}_\nu
-(\frac12-\alpha+6\beta)RR_{\mu\nu}-(1-\alpha+4\beta)R^{\sigma\rho}R_{\sigma\rho}
g_{\mu\nu}\nonumber \\
+\,(\frac{5}{16}-\frac{\alpha}2+2\beta)R^2g_{\mu\nu}+\frac{q}2\,\lambda^2g_{\mu\nu}=0\,\,
.\,\, \label{eqc10}
\end{eqnarray}
Immediately, the consistence with (A)dS solutions is evaluated by
replacing the contractions of $R_{\rho\mu\nu\sigma}= \lambda
(g_{\rho\sigma}g_{\mu\nu}-g_{\rho\nu}g_{\mu\sigma})$  in
(\ref{eqc10}). This gives
\begin{eqnarray}
q(\alpha)=\frac32-4\alpha\,\, ,\,\, \label{eqc10a}
\end{eqnarray}
and this indicates that if $\alpha=\frac38$ is introduced in
action (\ref{eqc5}) get implicit (A)dS solutions from its field
equations.

Now we take a look on the $GTMG\lambda$ formulation, considering
the YM-extended action at the torsionless limit, this means
\begin{eqnarray}
S'={S^{(3)}}_{(\alpha,\beta)}+\frac{m\kappa^2}{2}\big<\varepsilon^{\mu\nu\lambda}\,tr\big(
A_\mu\partial_\nu A_\lambda +\frac{2}{3}\, A_\mu A_\nu A_\lambda
\big)\big>+\kappa^{2} \big<q(\alpha)\lambda^2\big>\nonumber
\\ +\kappa^2\big<b_{\alpha\beta}\,\varepsilon^{\beta\lambda\sigma}
{(A_\lambda)^\alpha}_\sigma \big>\,\, ,\,\, \label{eqc11}
\end{eqnarray}
where $q(\alpha)$ is defined by (\ref{eqc10a}) then, this action
is consistent with (A)dS solutions when $m=0$. Variations on the
metric conduce to the known equations (\ref{eqc10}). So, the
connection field equation is
\begin{eqnarray}
\nabla_\mu {(\mathcal{F}^{\mu\nu})^\sigma}_\rho+
\frac{m}2\,\varepsilon^{\alpha\beta\nu}{(F_{\alpha\beta})^\sigma}_\rho
+b_{\rho\mu}\,\varepsilon^{\mu\nu\sigma}=0\,\, ,\,\, \label{eqc12}
\end{eqnarray}
and ${(\mathcal{F}^{\mu\nu})^\sigma}_\rho$ is defined in
(\ref{eqc5b}). Recalling that
$(F_{\mu\nu})_{\alpha\beta}=R_{\alpha\beta\nu\mu}$ in a
torsionless space-time, equation (\ref{eqc12}) can be rewritten in
terms of Ricci's tensor as follows
\begin{eqnarray}
(\frac12-4\beta)\nabla_\rho
R_{\nu\sigma}-(\frac32+4\beta)\nabla_\sigma
R_{\nu\rho}+(\frac12+2\beta)g_{\nu\rho}\partial_\sigma R +2\beta
g_{\nu\sigma}\partial_\rho R\nonumber
\\-\,m{\varepsilon^{\alpha\beta}}_\nu
(g_{\alpha\sigma}R_{\beta\rho}-g_{\alpha\rho}R_{\beta\sigma}-\frac{R}{2}\,
g_{\alpha\sigma}g_{\beta\rho})+
b_{\rho\mu}\,{\varepsilon^\mu}_{\nu\sigma}=0\,\, .\,\,
\label{eqc13}
\end{eqnarray}
Performing some algebraic manipulation on this last equation,
conditions (\ref{eqc8a}) and (\ref{eqc8c}), which establish the
symmetry property of Lagrange multipliers and the indetermination
of scalar curvature when $\beta=-\frac18$, rise again in a similar
way that they do in the massless theory.

Then, using condition (\ref{eqc9}), the Lagrange multipliers are
given by
\begin{eqnarray}
b_{\mu\nu}=2\bigg(\frac{\beta+\frac18}{\beta-\frac58}\bigg)m
R_{\mu\nu}-\bigg(\frac{\beta+\frac38}{\beta-\frac58}\bigg)\frac{mR}2
g_{\mu\nu}\,\, ,\,\, \label{eqc14}
\end{eqnarray}
and if (\ref{eqc7a}) is fixed, the  result (\ref{eqa14a}) is
recovered. So, evaluating the theory on $\beta=-\frac18$, the
action  (\ref{eqc13}) becomes in a similar form as in
(\ref{eq13}), this means
\begin{eqnarray}
\nabla_\mu \mathcal{R}_{\sigma\lambda}-\nabla_\lambda
\mathcal{R}_{\sigma\mu}
-m\,{\varepsilon^{\nu\rho}}_\sigma(g_{\lambda\nu}\mathcal{R}_{\mu\rho}-g_{\mu\nu}\mathcal{R}_{\lambda\rho}
-\frac23\,\mathcal{R}\,g_{\lambda\nu}g_{\mu\rho}) =0\,\, ,\,\,
\label{eqc15}
\end{eqnarray}
where again $\mathcal{R}_{\mu\nu}$ is defined as in (\ref{eqc9b})
and the trace $\sigma-\lambda$ is an identity, as one can expect.
\vskip .2truein
\section{Concluding remarks}

A perturbative regime based on arbitrary variations of the
contorsion and metric as a (classical) fixed background, is
performed in the context of a pure Yang-Mills formulation of the
$GL(3,R)$ gauge group. There, we analyze in detail the physical
content and the well known fact that a variational principle based
on the propagation of torsion (contorsion), as dynamical and
possible candidate for a quantum canonical description of gravity
in a pure YM formulation gets serious difficulties.

In the $2+1$ dimensional massless case we show that the theory
propagates three massless degrees of freedom, one of them a
non-unitary mode. Then, introducing appropiate quadratical terms
dependent on torsion, which preserve parity and general
covariance, we can see that the linearized limit do not reproduces
an equivalent pure Hilbert-Einstein-Fierz-Pauli massive theory for
a spin-2 mode and, moreover there is other non-unitary modes.
Roughly speaking, at first sight one can blame it on the kinetic
part of YM formulation because the existence of non-positive
Hamiltonian connected with non-unitarity problem. Nevertheless
there are other possible models of Gauss-Bonnet type which could
solve the unitarity problem.

Exploring the massless and the topological massive gravity models
in $2+1$ dimension, the well known existence of a YM-extended
theories family is noted. This family is labeled with two free
parameters, $\alpha$ and $\beta$ and can cure non-unitary
propagations.

Nevertheless, when the classical consistence between these type of
theories and the Einstein's one is tackled, what we have mentioned
as {\it torsionless limit}, it is shown that the parameter
$\alpha$ is related with the coupling of the cosmological constant
in the action.

Meanwhile, the parameter $\beta$ get two types of critical values.
On one side, the number $\beta=\frac58$ is connected to the
classical consistence requirement which demands the introduction
of torsion's Lagrangian constraints with solvable Lagrange
multipliers. On the other side, the value $\beta=-\frac18$
establishes a wide set of theories, including the Einstein's
solutions after the imposition of a auxiliary condition
$R=constant$ and non-Einsteinian ones when the Ricci scalar became
an arbitrary function. But, even though the Lagrangian extension
of the YM formulation for gravity conduces to the well known fact
that there exists unphysical classical solutions, the same occurs
(in a much less severe way) without these corrections and one can
recall the YM pure formulation gives rise a set of solutions for
the massless and topological massive gravity with the property
$R=constant$ and only Einsteinian results can be obtained if the
auxiliary condition $R=-6\lambda$ is fixed.

A picture with a little bit of generalization including a dynamic
metric and non-Minkowskian background in the perturbative analysis
would be considered elsewhere.





\end{document}